\documentclass[useAMS,usenatbib,usegraphicx]{mn2e}

\pagerange{\pageref{firstpage}--\pageref{lastpage}}
\pubyear{2010}

\newcommand{\be}{\begin{equation}}
\newcommand{\ee}{\end{equation}}

\begin{document}

\title[Power Asymmetries in the Cosmic Microwave Background Temperature and Polarization patterns]
{Power Asymmetries in the Cosmic Microwave Background Temperature and Polarization 
patterns}

\author[F. Paci, A. Gruppuso, F. Finelli, P. Cabella, A. De Rosa, N. Mandolesi, P. Natoli]
{F.~Paci$^{1}$\thanks{E-mail: paci@ifca.unican.es}, A.~Gruppuso$^{2,3}$, F.~Finelli$^{2,3}$, P.~Cabella$^{4}$, A.~De Rosa$^{2}$,
N.~Mandolesi$^{2}$, 
\newauthor P.~Natoli$^{4,5}$ \\
$^1$ 
Instituto de F\'isica de Cantabria (CSIC - Univ. de Cantabria), Avda. Los Castros s/n, 39005 Santander, Spain\\
$^2$ 
INAF-IASF Bologna, Istituto di Astrofisica Spaziale e Fisica Cosmica 
di Bologna, via Gobetti 101, I-40129 Bologna, Italy \\
$^3$
INFN, Sezione di Bologna, Via Irnerio 46, I-40126 Bologna, Italy \\
$^4$
Dipartimento di Fisica, Universit\'a di Roma Tor Vergata, Via della Ricerca Scientifica 1, I-00133 Roma, Italy \\
$^5$
INFN, Sezione di Tor Vergata, Via della Ricerca Scientifica 1, I-00133 Roma, Italy }

\label{firstpage}

\maketitle

\begin{abstract}
We test the asymmetry of the Cosmic Microwave Background anisotropy
{\em jointly in temperature and polarization}. We study the hemispherical
asymmetry, previously found only in the temperature field, with respect to the axis
identified by Hansen et al. (2009). 
To this extent, we make use of the low resolution WMAP 5 year temperature
and polarization $N_{\rm side}=16$ maps and the optimal angular power
spectrum estimator {\it BolPol} \citep{Gruppuso2009}.
We consider two simple estimators for the power asymmetry
and we compare our findings with Monte Carlo simulations which take into
account the full noise covariance matrix.
We confirm an excess of power in temperature angular power spectrum
in the Southern hemisphere at a
significant level, between $3 \sigma$ and $4 \sigma$ depending
on the exact range of multipoles considered.
We do not find significant power asymmetry in the gradient (curl)
component $EE$ ($BB$) of polarized angular spectra.
Also cross-correlation power
spectra, i.e. $TE$, $TB$, $EB$, show no significant hemispherical asymmetry.
We also show that the Cold Spot found by \citet{Vielva2004} in the Southern
Galactic hemisphere does not alter the significance of the hemispherical
asymmetries on multipoles which can be probed by maps at resolution $N_{\rm side}=16$.
Although the origin of the hemispherical asymmetry in temperature remains unclear, 
the study of the polarization patter could add useful information on its explanation.  
We therefore forecast by Monte Carlo the {\sc Planck} capabilities in probing polarization asymmetries.
\end{abstract}


\section{Introduction}

Great attention has been devoted to a hemispherical
power asymmetry in the intensity pattern of the Cosmic Microwave Background
(CMB) as seen by WMAP \citep{Hinshaw:2008kr,dunkley_wmap5}.
Such asymmetry has been originally found in WMAP 1st year release and appears to lay on
an axis nearly orthogonal to the ecliptic plane \citep{Eriksen:2003db,Hansen:2004vq}. It 
has been confirmed in the WMAP three year and five year release \citep{Eriksen:2007pc,Hansen:2008ym,Hoftuft:2009rq} 
and it is present in the COBE data as well, although with lower significance. The temperature power spectra
of the opposing hemispheres are inconsistent at $3 \sigma$
to $4 \sigma$ depending on the range of multipoles considered.
The asymmetry has been detected in low resolution maps \citep{Eriksen:2003db}, both in
angular and multipoles space, but it extends to much smaller angular scales in
the multipole range $\delta \ell = [2,600]$ \citep{Hansen:2008ym}.
It is unclear whether this hemispherical
asymmetry is primordial or due to unknown residual foreground/systematics.

Whereas several groups have performed different and independent
investigations on the CMB temperature pattern, the joint analysis of the
CMB {\em temperature and polarization} pattern has not been performed yet.
The information encoded in the polarization pattern may turn out extremely
useful to clarify the presence of the hemispherical asymmetry shedding light on its origin.
Low resolution WMAP 5 year
maps in $(T,Q,U)$ with relative noise covariance matrices
are publicly available: these public maps have allowed a re-analysis by a
quadratic maximum likelihood (henceforth QML) estimator of the
low multipole angular power spectrum in temperature and polarization
\citep{Gruppuso2009}.

In this paper we address the issue of hemispherical
asymmetry by estimating the power spectrum in the two hemispheres by using the QML: our application of QML
in this context is novel and extremely useful since the
aggressive masking needed to reduce residual foreground contamination might be even more problematic for polarization
than for temperature \citep{Bunn:2002df,Smith:2006vq}.

Our main aim is to
test whether other asymmetries in full temperature-polarization pattern are
present around the most recently determined axis defined by the direction
$( \theta=107 , \phi=226)$ \citep{Hansen:2008ym}, where $\theta$ and $\phi$ are the Galactic
colatitude and longitude, respectively.

This paper is organized as follows. In
Section 2 we describe our
methodology by reviewing
the algebra of the QML estimator. We also discuss the data set used
and introduce the $R$ and $D$ estimators, the ratio and the difference of the power in the two hemispheres respectively.
In Section 3 we discuss our
results including the related Monte Carlo uncertenties
based on 1000 simulations. We discuss {\sc Planck} predicted performances in
probing the hemispherical asymmetries in Section 5, while in Section 6 we draw
our main conclusions.

\section{Description of the analysis}

\subsection{Angular Power Spectra Estimation}

In order to evaluate the angular power spectra, we use the {\it BolPol} code, a QML estimator.
The QML formalism was introduced in \citet{tegmark_tt} and extended to polarization in \citet{tegmark_pol}. 
In this section we describe the essence of the method. Further details
can be found in \citet{Gruppuso2009} where {\it BolPol}) 
has been applied to WMAP 5 year low resolution data.

Given a map in temperature and polarization ${\bf x=(T,Q,U)}$, the QML provides estimates
$\hat {C}_\ell^X$ - with $X$ being one of $TT$, $EE$, $TE$, $BB$,
$TB$, $EB$ - of the angular power spectrum as: 
\be
\hat{C}_\ell^X = \sum_{\ell' \,, X'} (F^{-1})^{X \, X'}_{\ell\ell'} \left[ {\bf x}^t
{\bf E}^{\ell'}_{X'} {\bf x}-tr({\bf N}{\bf
E}^{\ell'}_{X'}) \right]
\, ,
\ee
where the Fisher matrix $F_{X X'}^{\ell \ell '}$ is defined as
\be
\label{eq:fisher}
F^{\ell\ell'}_{X X'}=\frac{1}{2}tr\Big[{\bf C}^{-1}\frac{\partial
{\bf C}}{\partial
  C_\ell^X}{\bf C}^{-1}\frac{\partial {\bf C}}{\partial
C_{\ell'}^{X'}}\Big] \,,
\ee
and the ${\bf E}^{\ell}_X$ matrix is given by
\be
\label{eq:Elle}
{\bf E}^\ell_X=\frac{1}{2}{\bf C}^{-1}\frac{\partial {\bf C}}{\partial
  C_\ell^X}{\bf C}^{-1} \, ,
\ee
with ${\bf C} ={\bf S}(C_{\ell}^X)+{\bf N}$ being the global covariance matrix (signal plus noise
 contribution\footnote{Note that, in principle it is possible to include in this matrix residuals from foreground subtraction.
This is the case for the WMAP foreground reduced covariance matrix we employ hereafter}) and $C_{\ell}^X$ is a fiducial power spectrum. 

Although an initial assumption for a fiducial power spectrum
$C_\ell^X$ is needed, the QML method provides unbiased estimates
of the power spectrum contained in the map regardless of the initial guess,
\be
\langle\hat{C}_\ell^X\rangle=\bar{C}_\ell^{X} \,,
\label{unbiased}
\ee
where the average is taken over the ensemble of realizations (or, in a practical test, over Monte Carlo realizations extracted from $\bar{C}_\ell^{X}$).

On the other hand, the covariance matrix associated to the estimates,
\be
\langle\Delta\hat{C}_\ell^X
\Delta\hat{C}_{\ell'}^{X'} \rangle= ( F^{-1})^{X \, X'}_{\ell\ell'} \,,
\label{minimum}
\ee
does depend on the assumption for the fiducial power spectrum $C_\ell^X$: 
the closer the guess to the true power spectrum is, the closer are the error bars to minimum variance. 
According to the Cramer-Rao inequality, Eq. (\ref{minimum}) tells us that 
the QML has the smallest error bars. 
We thus call the QML an `optimal' estimator.

\subsection{Data set and Simulations}

In this Section we describe the data set that we have considered and the corresponding simulations we have produced to analyze it. 
We use the temperature ILC map smoothed at $9.8$ degrees and reconstructed at HealPix\footnote{http://healpix.jpl.nasa.gov/}
 \citep{gorski} resolution $N_{\rm side}=16$, 
the foreground cleaned low resolution maps and the noise covariance matrix in $(Q,U)$ publicly available at the LAMBDA website\footnote{http://lambda.gsfc.nasa.gov/}. 
We have added to the temperature map a random noise realization with variance of $1 \mu K^2$, as suggested in \citet{dunkley_wmap5}.
Consistently, the noise covariance matrix for $TT$ is taken to be diagonal with variance equal to $1 \mu K^2$.

\begin{figure}
\includegraphics[scale=0.32, angle=90]{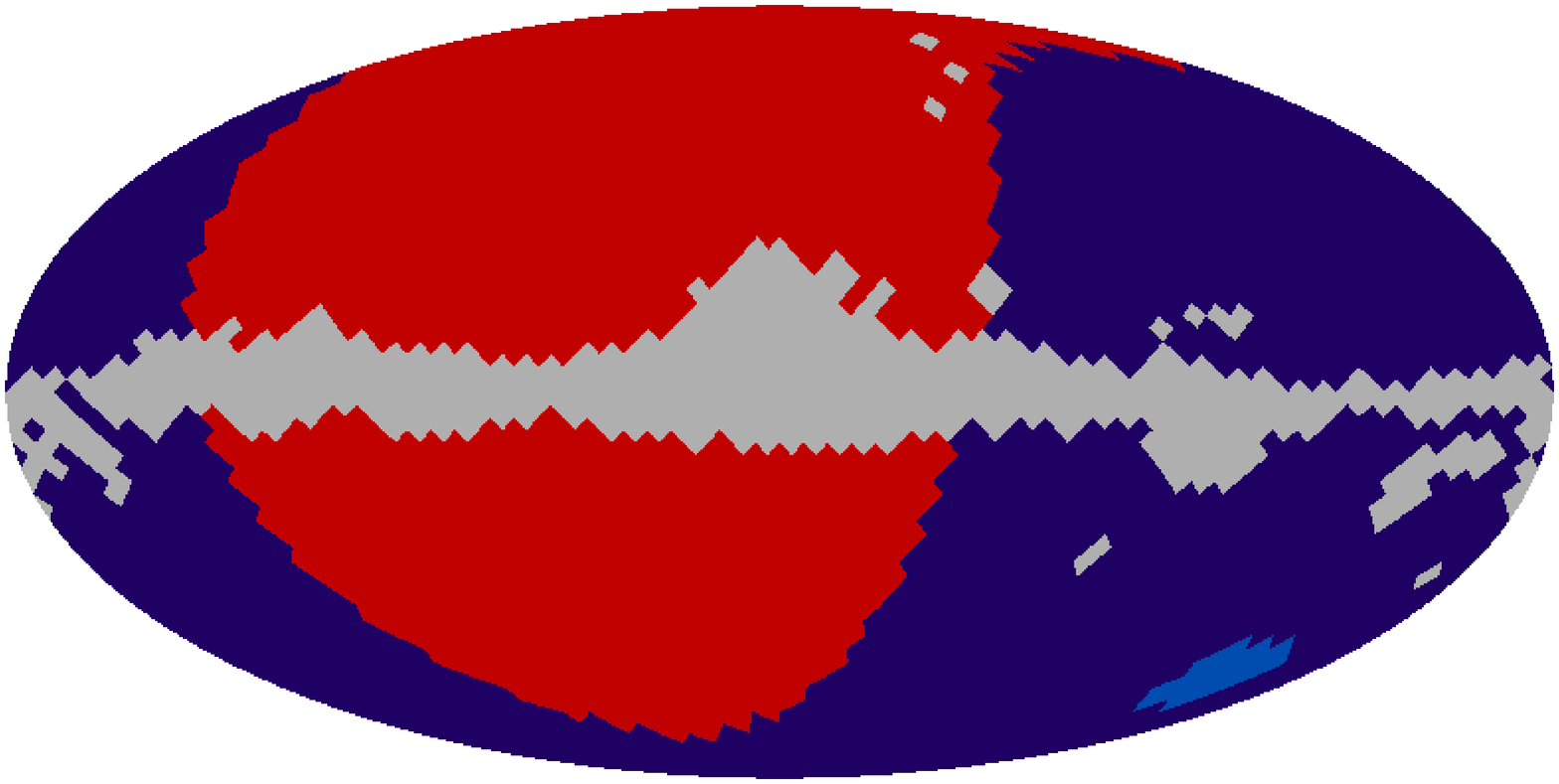}
\includegraphics[scale=0.32, angle=90]{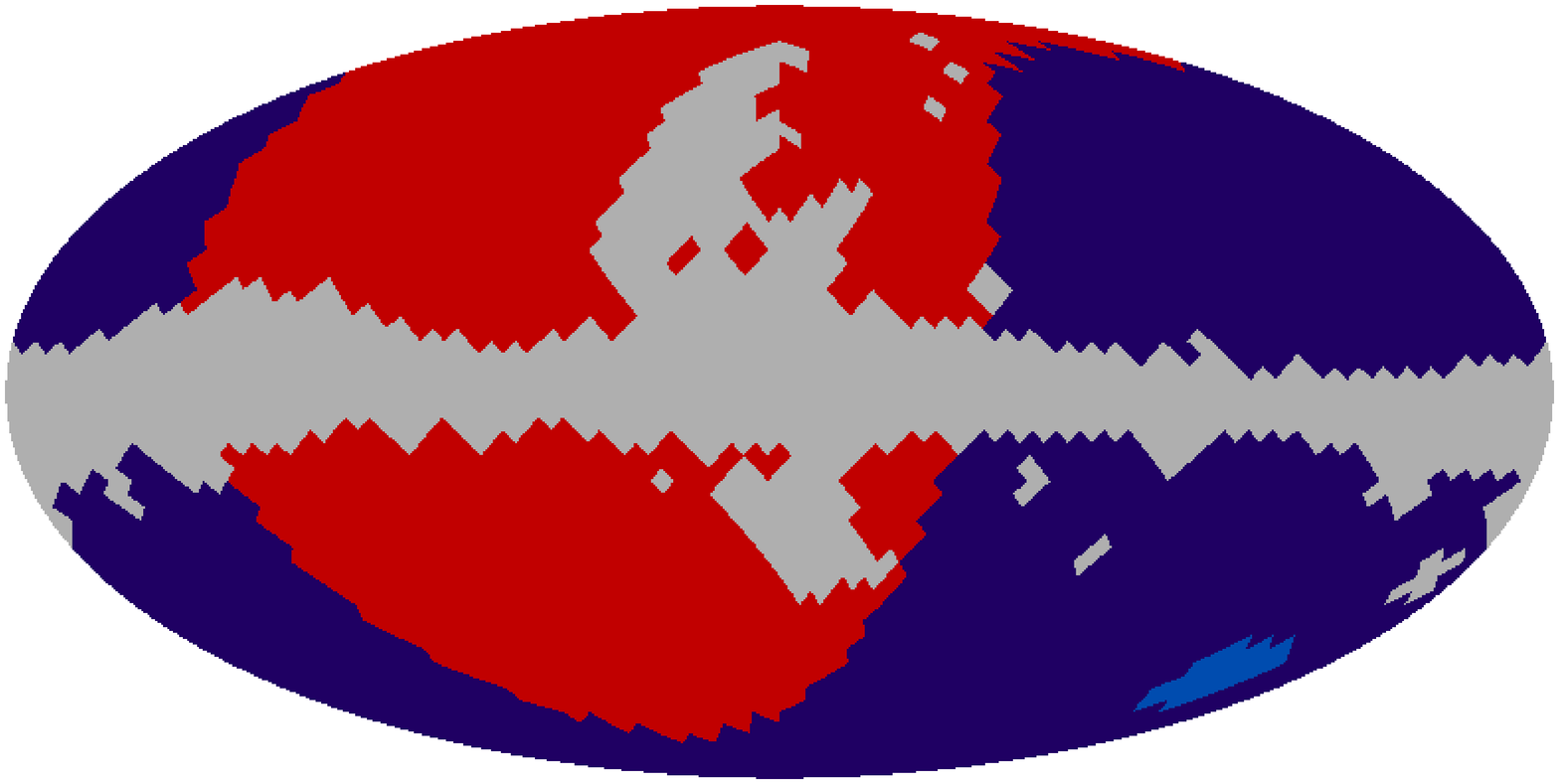}
\caption{Mollweide projection of the observed Northern (red) and Southern (blue) hemisphere at $N_{\rm side}=16$.
The (light blue) circle in the Southern hemisphere corresponds to the region of the Cold Spot, whereas the light grey region
corresponds to the WMAP low resolution galactic mask for temperature (upper panel) and polarization (lower panel).
}
\label{masks}
\end{figure}

To perform the analysis, we have built the masks for the two hemispheres defined by 
the direction $( \theta=107 , \phi=226)$ \citep{Hansen:2008ym} and combined them
with the Galactic WMAP 5yr low resolution temperature and polarization mask. 
Maps and covariances for the two sky regions (namely North and South) have been consistently tailored to the produced masks
(see Figure \ref{masks}).

To assess the significance of the power asymmetries found in the data, our results have been tested against Monte Carlo simulations. 
A set of 1000 CMB+noise sky realizations has been generated: the signal was generated from the WMAP 5 year best fit model, the noise through 
a Cholesky decomposition of the global ($T,Q,U$) noise
covariance matrix. We then computed the angular power spectra for each of the 1000 simulations using {\it BolPol} and built two figures of merit as explained in the next subsection.

\subsection{Estimators}

We define the following quantities
\be
C^{X}_{N/S} \equiv {1\over {(\ell_{max}-1)}} \, \sum_{\ell=2,\ell_{max}} {\ell (\ell + 1) \over{2 \pi }} \, \hat{C}^{X, N/S}_{\ell} 
\label{CNS}
\ee
where $\hat{C}^{X, N}_{\ell}$ and $\hat{C}^{X, S}_{\ell}$ are the estimated angular power spectra obtained with {\it BolPol} observing only the Northern (`$N$') and the Southern (`$S$') hemisphere respectively, outside the galactic plane. 
As above, $X$ runs over the spectral types. 

Two estimators can be built as follows: 
the ratio $R^X$, as performed in \citet{Eriksen:2003db},
\be
R^X = C^X_S/C^X_N \, ,
\label{rapporto}
\ee
and the difference $D^X$,
\be
D^X=C^X_S - C^X_N \, ,
\label{differenza}
\ee
of the two aforementioned quantities. 
In the following, we will drop the index $X$ for $R$ and $D$ specifying only the spectrum we refer to.
 
For our application to WMAP data, both estimators have been considered for $TT$, while only the $D$ estimator 
has been applied to the other spectra ($EE$, $TE$, $BB$, $TB$ and $EB$), because of unfavorable 
signal-to-noise ratio of the WMAP data in polarization.

\section{Results}

\begin{figure*}
\includegraphics[width=15cm,height=10cm,angle=0]{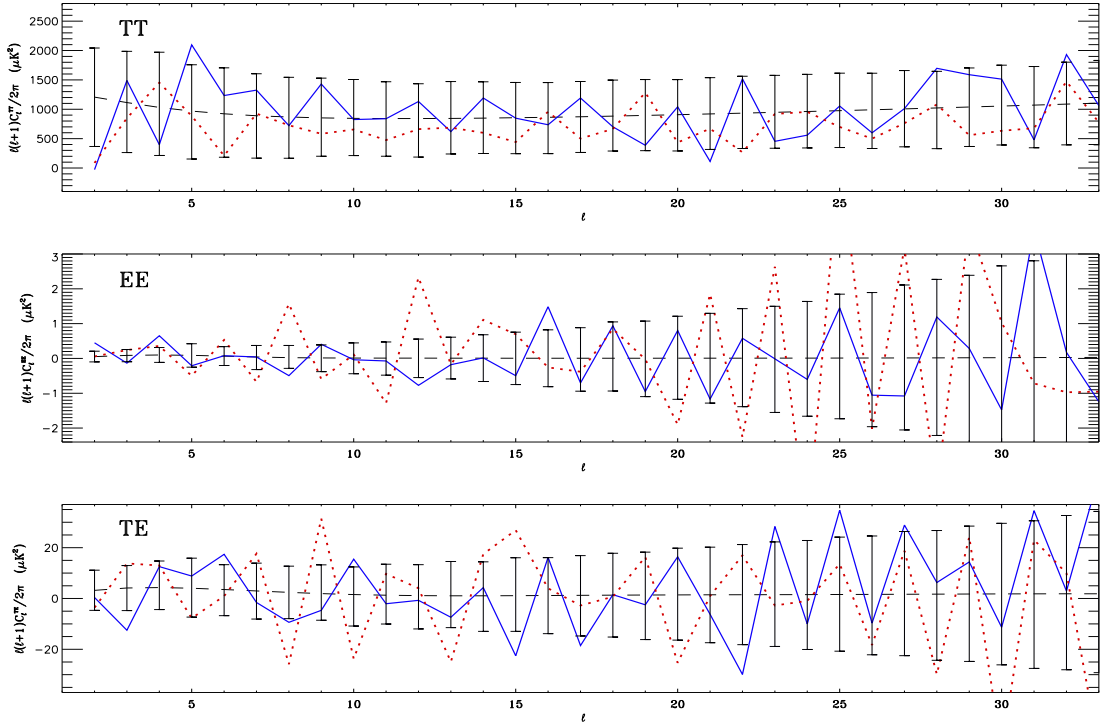}
\caption{QML estimates for $TT$ (upper panel), $EE$ (middle panel) and $TE$ (lower panel) from WMAP 5 year 
$N_{\rm side}=16$ maps. Solid (blue) line is for the angular power spectrum of the Southern hemisphere (blue region of
Fig. \ref{masks}), whereas dotted (red) line is for the Northern one (red region of Fig. \ref{masks}).
Dashed line shows the WMAP 5 year best fit, taken as fiducial power spectrum for the analysis. For reference, 
we also show the error bars of the QML computed from a Monte Carlo of 1000 sky realizations of the Northern hemisphere with the global ($T$,$Q$,$U$) noise covariance matrix 
(error bars from the Monte Carlo on the Southern hemisphere are basically undistinguishable from the ones plotted).}
\label{estimates1}
\end{figure*}

\begin{figure*}
\includegraphics[width=15cm,height=10cm,angle=0]{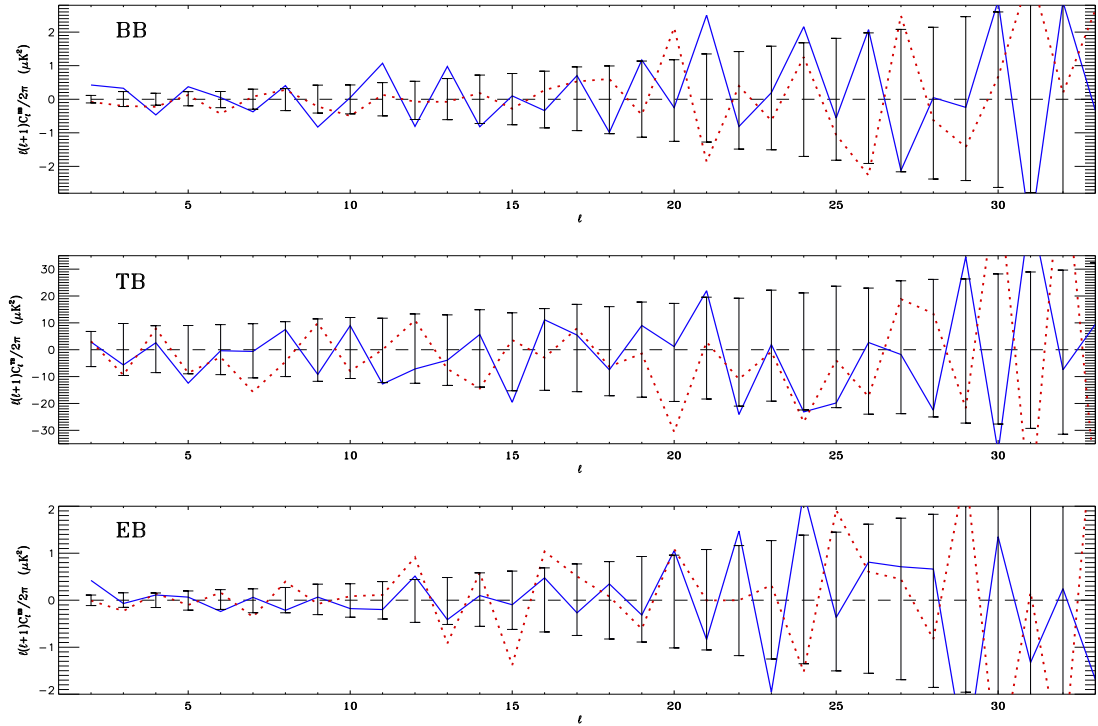}
\label{estimates2}
\caption{As in Fig. \ref{estimates1} but for $BB$, $TB$, $EB$ (from top to bottom)}
\end{figure*}

The six angular power spectra $TT$, $EE$, $TE$, $BB$, $TB$, $EB$ are presented in Fig. \ref{estimates1} 
and \ref{estimates2}. Our results for $TT$, shown in the upper panel of Fig.~\ref{estimates1}, are  
consistent with those obtained by \citet{Eriksen:2003db}.

\begin{figure}
\includegraphics[width=8.0cm,angle=0]{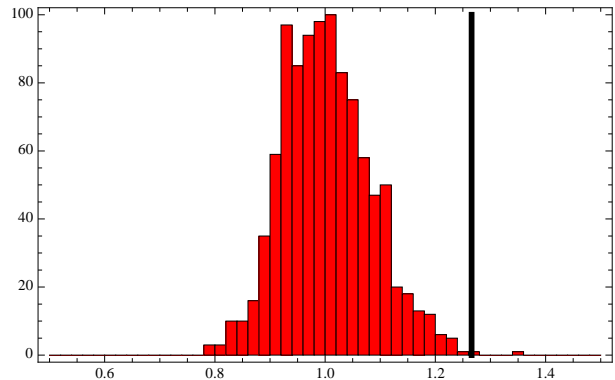}
\caption{$TT$. Number counts ($y$-axis) versus $R$ ($x$-axis) for the range $\Delta \ell = [2,40]$. The vertical line stands for  the WMAP 5 yr data. 
The probability to obtain a smaller value than the WMAP one is $99.8\%$.}
\label{RdistribTT}
\end{figure}

In Fig. \ref{RdistribTT} we show the $R$ estimator distribution for the range $\ell=2-40$.  For this estimator we obtain that the probability of having 
the WMAP value is as low as $0.2 \%$, which agrees with the results by \citet{Eriksen:2003db}.
In Table \ref{tableprobabilities1} the probability of obtaining the WMAP value for the $D$ estimator 
is computed for the following four multipoles ranges: $2-8$, $2-16$, $2-32$ and $2-40$. See Fig.~5 for the full empirical (Monte Carlo) probability distribution functions.
Note that the $R$ and $D$ estimator detect a comparable level of anomaly in the multipole range $2-40$.

In Table \ref{tableprobabilities2} we provide results for polarization and cross-spectra. As mentioned above, only $D$  is considered and computed for the four aforementioned multipoles range.
The estimator $R$, in fact, is not well defined any time the denominator $C_N^X$ approaches to zero, which might be the case for highly noisy spectra.
Although Table \ref{tableprobabilities2} does not show any significant deviation from the symmetry for polarization and
cross-spectra, it is nonetheless worth noting the behaviour of $EE$ in the range $\ell=2-16$, for which the probability of having the WMAP value is as low as $3.5 \%$, and of $BB$ in the range $\ell=2-8$ where the probability decreases to $2.2\%$. Moreover, an unexpected statistics seems to show up for $TE$ in the range $2-40$ where the probability of having the WMAP value is $0.5\%$. However, this mainly comes from the multipoles between $32$ and $40$, which are close to the threshold of reliability of the QML on $N_{\rm side}=16$ maps.

\begin{table}
\caption{Probabilities (in percentage) to obtain a smaller value than WMAP low resolution data for $TT$ angular power spectrum and the $D$-estimator.} 
\centering 
\begin{tabular}{c c c c c} 
\hline\hline 
 $D$ &  $\Delta \ell$ = 2-8 &  $\Delta \ell$ = 2-16 & $\Delta \ell$ = 2-32 &  $\Delta \ell$ = 2-40 \\ [0.5ex] 
\hline 
$TT$ & 86.2 & 96.9 & 99.8 & 99.1 \\  [1ex] 
\hline 
\end{tabular}
\label{tableprobabilities1} 
\end{table}

\begin{table}
\caption{Probabilities (in percentage) to obtain a smaller value than WMAP low resolution data} 
\centering 
\begin{tabular}{c c c c c} 
\hline\hline 
 $D$ &  $\Delta \ell$ = 2-8 & $\Delta \ell$ = 2-16 &  $\Delta \ell$ = 2-32 & $\Delta \ell$ = 2-40 \\ [0.5ex] 
\hline 
$TE$ & 59.9 & 16.9 & 75.6 & 99.5  \\ 
$EE$ & 5.4 & 3.5 & 28.8 & 34.3 \\
$BB$ & 97.8 & 79.5 & 71.9 & 81.0 \\
$TB$ & 80.9 & 54.1 & 42.8 & 91.7 \\
$EB$ & 61.3 & 54.6 & 74.4 & 21.8 \\  [1ex] 
\hline 
\end{tabular}
\label{tableprobabilities2} 
\end{table}

\begin{figure}
\includegraphics[width=4.1cm,angle=0]{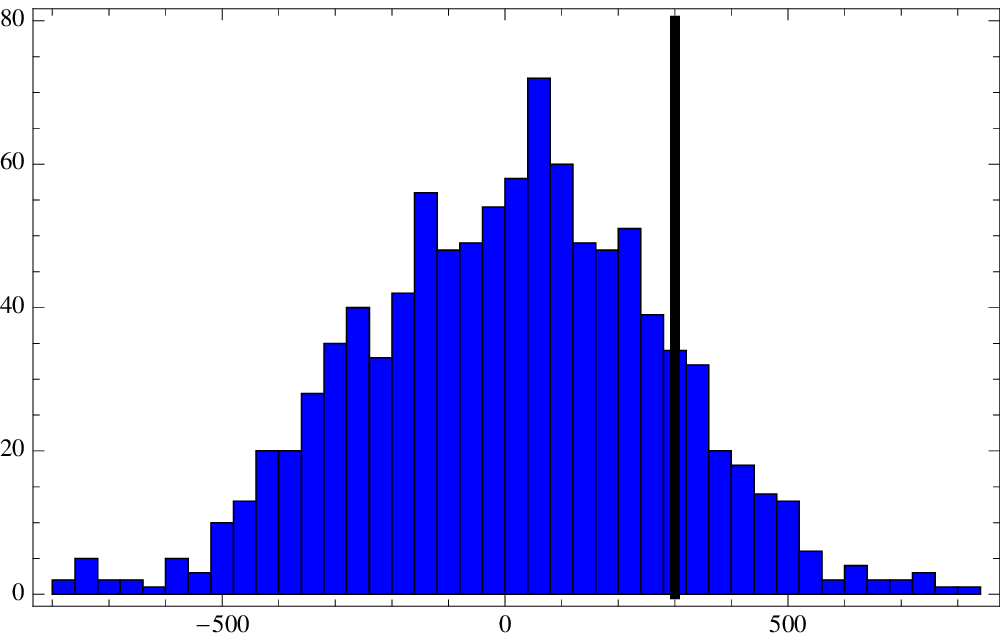}
\includegraphics[width=4.1cm,angle=0]{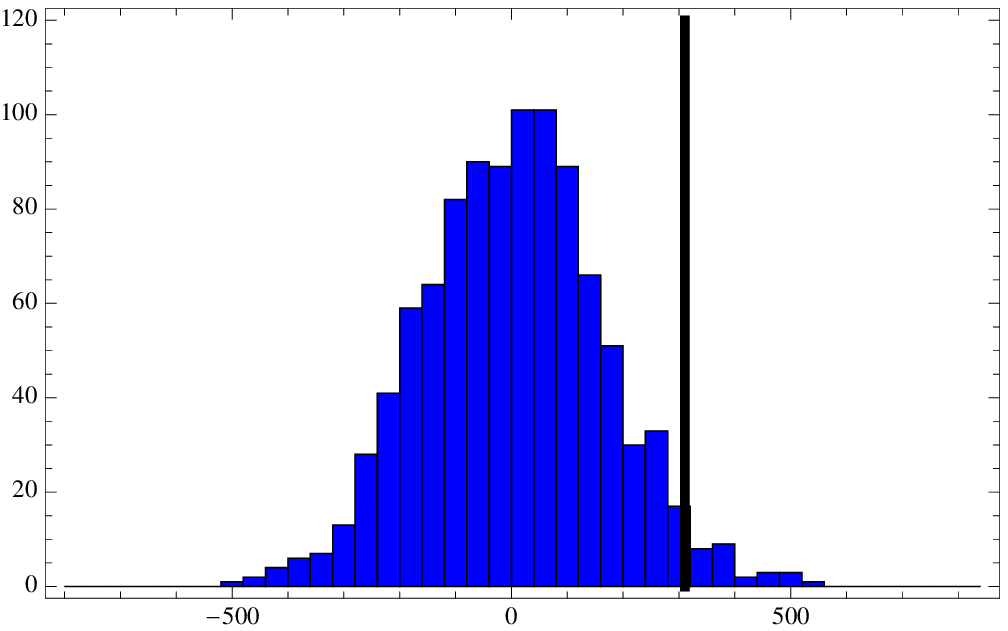}
\includegraphics[width=4.1cm,angle=0]{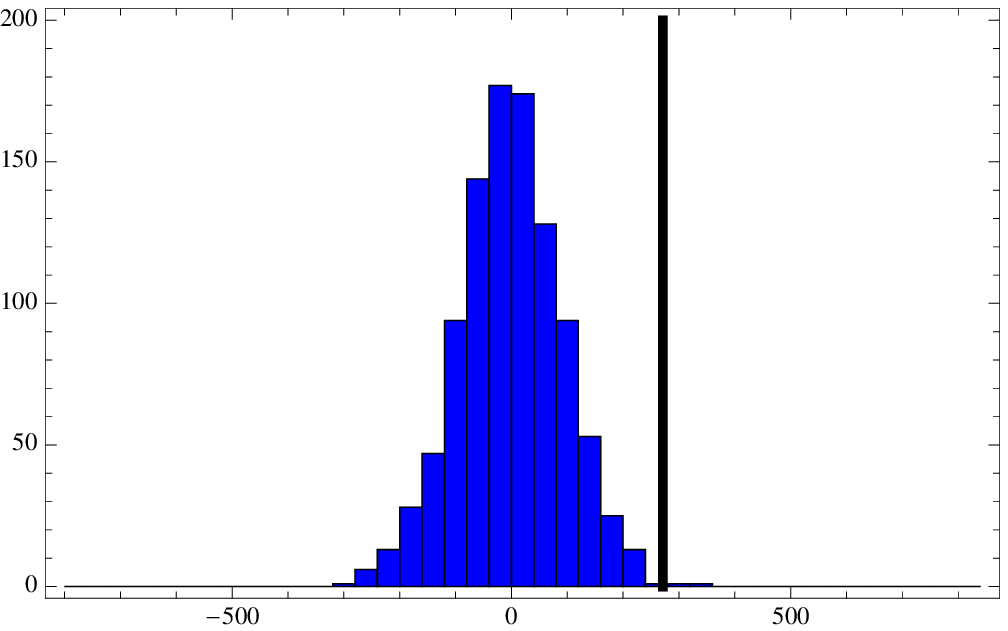}
\includegraphics[width=4.1cm,angle=0]{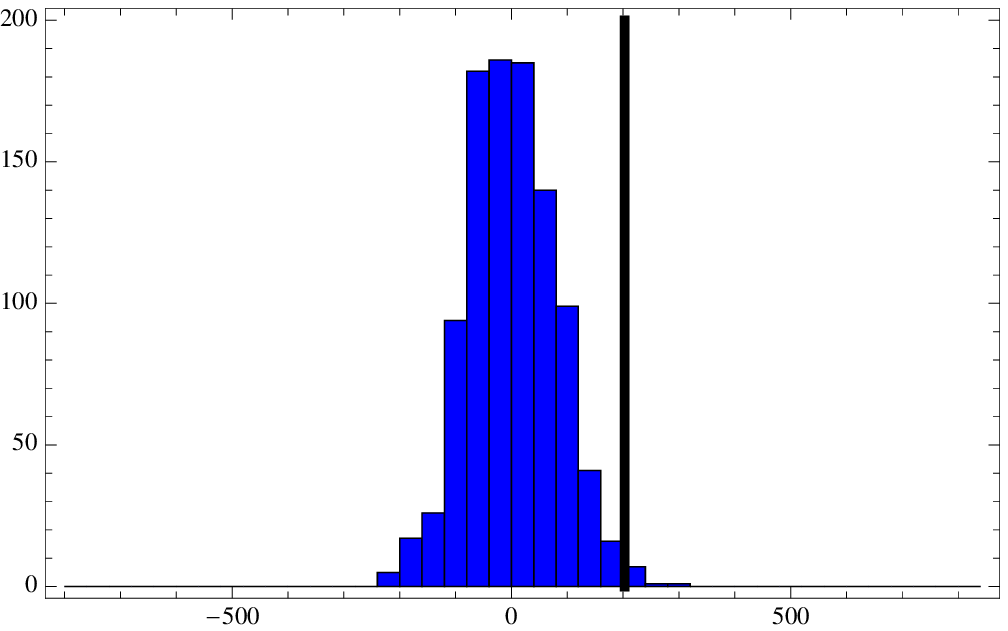}
\caption{$TT$. All the panels present the number counts ($y$-axis) versus the $D$ statistics ($x$-axis),
being the latter in units of $\mu K^2$. For each panel, $D$ has been computed within a different range of
multipoles. 
Top left: $\Delta \ell = [2,8]$. 
Top right:  $\Delta \ell = [2,16]$. Bottom left: $\Delta \ell = [2,32]$. 
Bottom right: $\Delta \ell = [2,40]$. Vertical lines for the WMAP 5 yr data.}
\label{DestimatorTT}
\end{figure}

\begin{figure}
\includegraphics[width=4.1cm,angle=0]{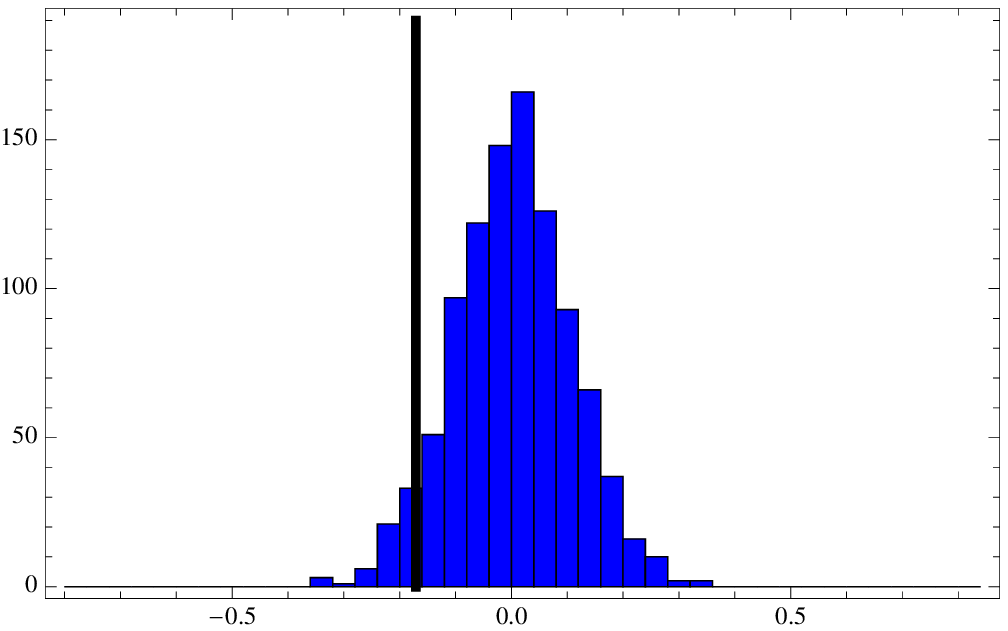}
\includegraphics[width=4.1cm,angle=0]{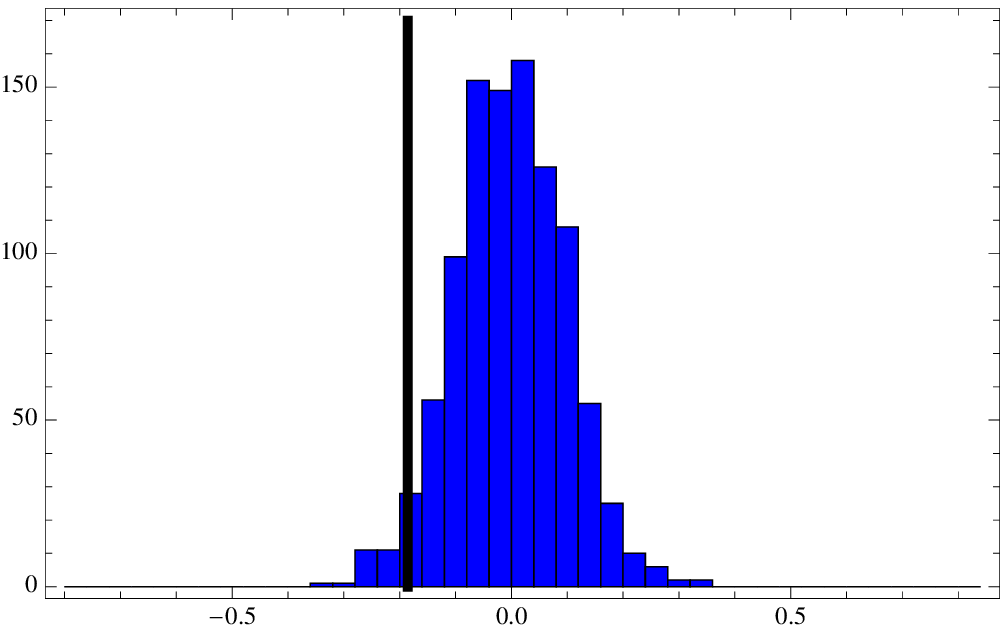}
\includegraphics[width=4.1cm,angle=0]{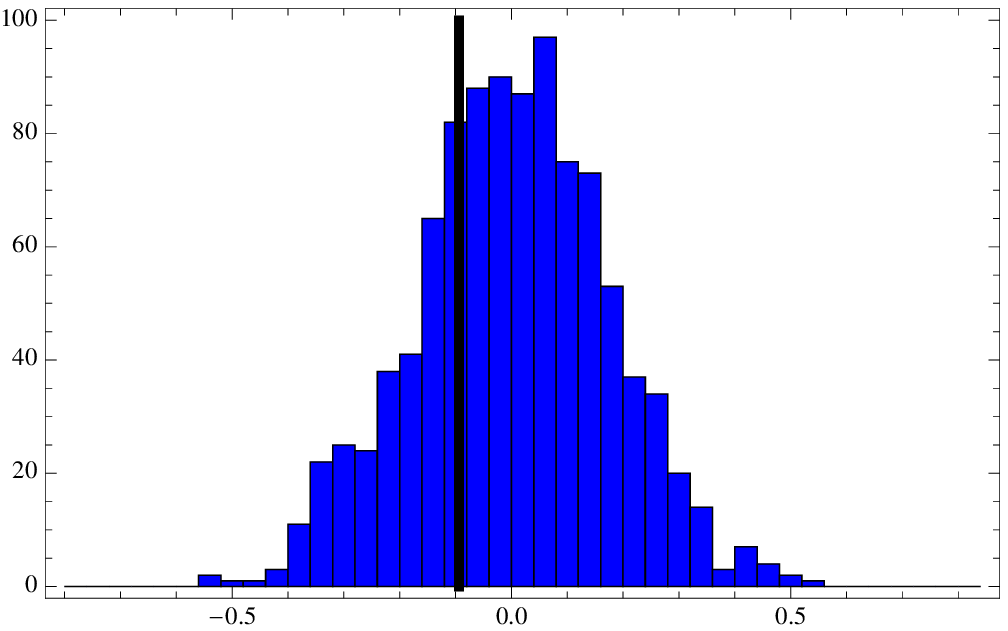}
\includegraphics[width=4.1cm,angle=0]{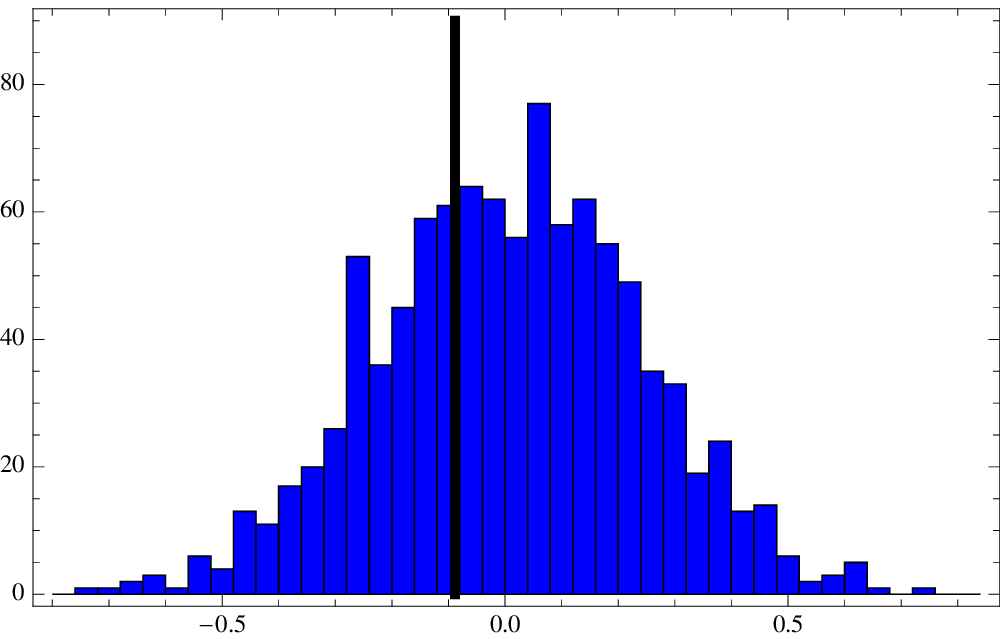}
\caption{As in Fig. \ref{DestimatorTT} but for $EE$.}
\label{DestimatorEE}
\end{figure}

We also report on the possible contribution to the North-South power asymmetry given by the Cold Spot found by \citet{Vielva2004} 
(see also \citet{Cruz2005, Cruz2007}). By masking out the Cold Spot (see light blue spot of Fig. \ref{masks}) with a circle of radius 8 degrees - a conservative choice compared to its size of 5 degrees - we have not found any significant deviation from the $C_\ell$ obtained without masking it out. This might be due to the fact that the low resolution of our data set prevents us from exploring properly the angular scales of interest for the physical size of the Cold Spot ($\ell \simeq 40$). Moreover, the smoothing process applied to the temperature map might be also responsible for washing out features like the Cold Spot.
Nonetheless, a possible connection between these large scale anomalies has been claimed recently in \citet{Bernui:2009pv}, where a different estimator with respect to the one adopted in the present analysis has been exploited. Withstanding all the caveats set forth above, our analysis suggests that the Cold Spot has little to do with WMAP 5 year asymmetries.

\section{Planck Forecasts}

The {\sc Planck} satellite \citep{bluebook} has been launched on May 14th, 2009, and it will measure CMB anisotropies 
with unprecedented precision.
In order to assess its capabilities in probing the hemispherical asymmetry, we consider the nominal
sensitivity of the {\sc Planck} 143 GHz channel, taken as representative of the results which can be obtained after the foreground 
cleaning from various frequency channels. The 143 GHZ channel has an angular resolution of $7.1'$ (FWHM) and 
an average sensitivity of $6 \mu K$ ($11.4 \mu K$) per pixel - a square whose side is the FWHM size of the beam - in temperature (polarization), after 2 full sky surveys \citep{bluebook}.
We assume uniform uncorrelated instrumental noise and 
we build the corresponding diagonal covariance matrix for temperature and polarization, 
from which, through Cholesky decomposition
we are able to extract noise realizations. 

As expected, we notice that no significant improvement will be achieved with {\sc Planck} for the $TT$ spectrum, since both WMAP and 
{\sc Planck} are cosmic variance limited for the range of multipoles considered here. 
On the contrary, polarization and cross-spectra do benefit from the {\sc Planck} increased sensitivity.
In Fig. 13 we plotted our forecasted distribution for the estimator $D$ for $TE$ and $EE$ 
on top of the same distribution for the WMAP case: the shrinking
of the distribution due to the higher sensitivity is more than evident.
Moreover, for these two cases we find that also the $R$ estimator yields valuable information.
Finally, we do not expect to be able to apply the $R$ estimator for any spectrum involving $B$ because of the low level of signal, if any. 
Distributions
of $D$ for $BB$, $TB$ and $EB$ are analogous to what shown for $TE$ and $EE$.

\begin{figure}
\includegraphics[width=4.1cm,angle=0]{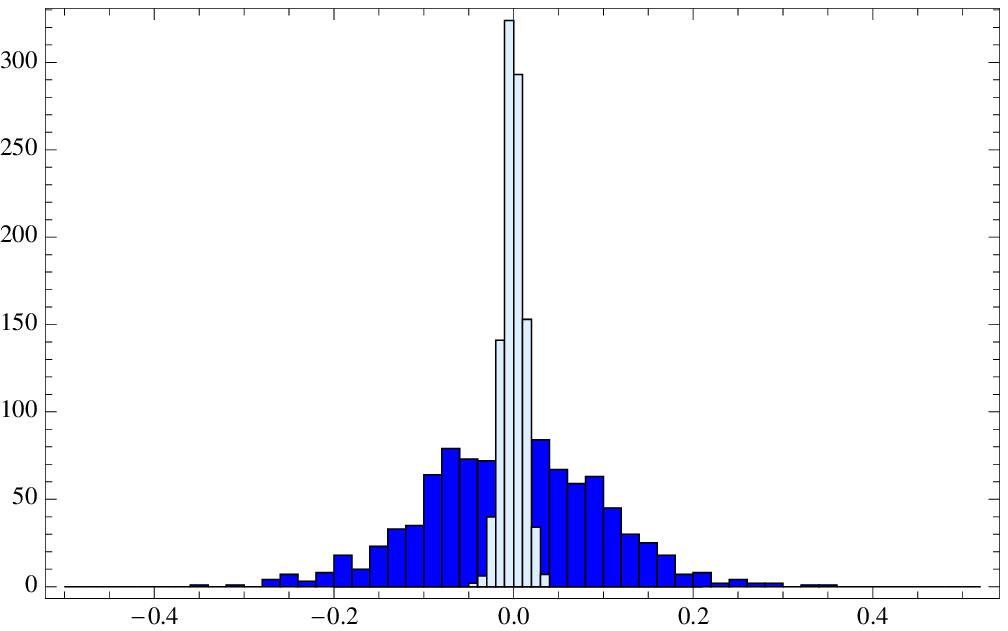}
\includegraphics[width=4.1cm,angle=0]{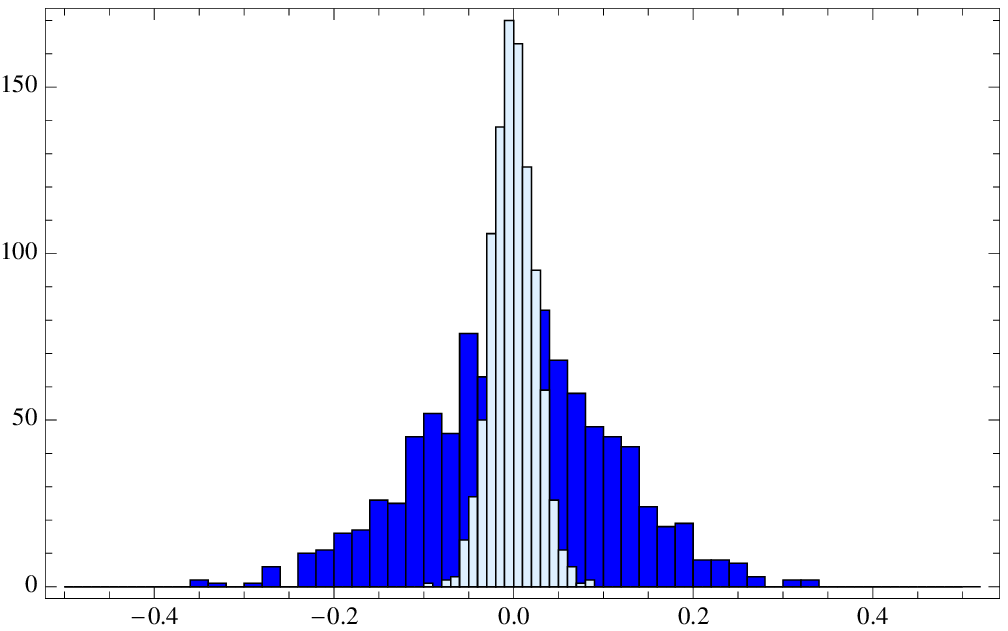}
\includegraphics[width=4.1cm,angle=0]{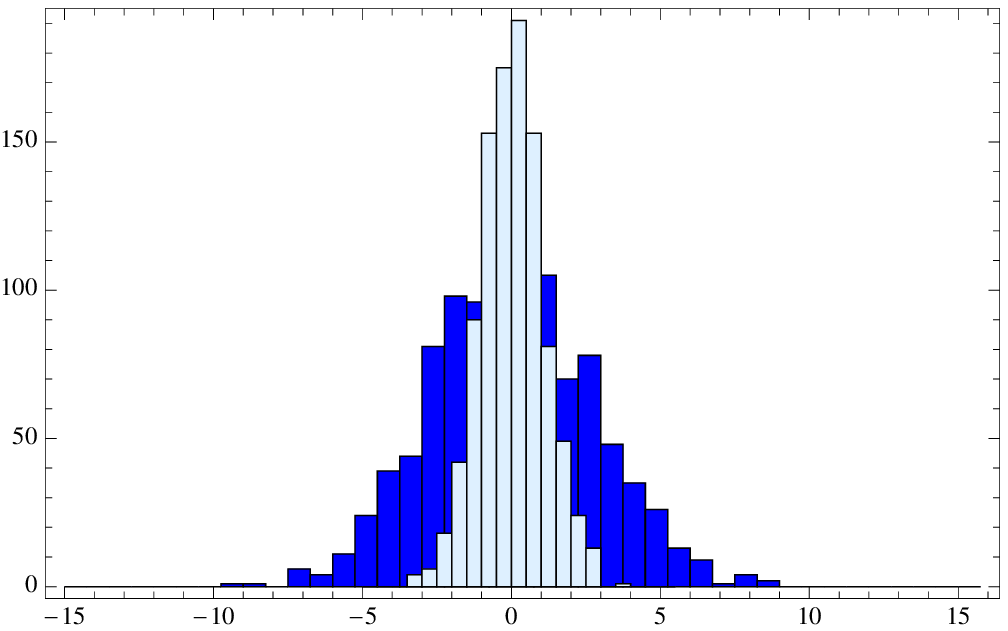}
\includegraphics[width=4.1cm,angle=0]{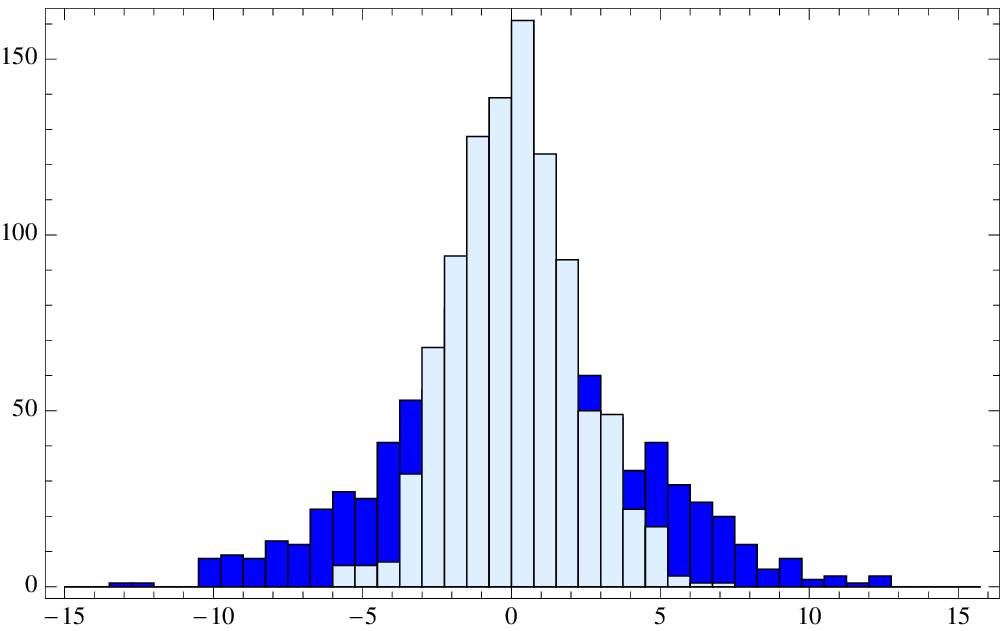}
\caption{Comparison between WMAP and {\sc Planck} for $EE$ (upper panels) and for TE (lower panels). Left panels: number counts for $D$ with $\Delta \ell = [2,16]$. 
Right panels: $D$ with $\Delta \ell = [2,8]$.
In all the panels, dark blue is for WMAP and light blue is for {\sc Planck}. Units for $D$ in the $x$-axis are $\mu K^2$.}
\label{compPlanckWMAP}
\end{figure}

\section{Discussions and Conclusions}

Using an optimal power spectrum estimator, we have confirmed the power asymmetry for $TT$ found by \citet{Eriksen:2003db}
along the direction reported in \citet{Hansen:2008ym}, see Table \ref{tableprobabilities1} and Fig.~5.
Considering the same axis, we have extended such analysis to the other spectra ($TE$, $EE$, $BB$, $TB$ and $EB$)
considering only the estimator $D$, defined in Eq.~(\ref{differenza}), because the noise level of WMAP permits the use of R 
(see Eq.~(\ref{rapporto})) only for $TT$.

Since our implementation of the QML \citep{Gruppuso2009} is capable of handling the full noise covariance matrix in ($T$,$Q$,$U$), the 
analysis of the present paper is joint for temperature and polarization. The information encoded in CMB polarization is 
complementary to the temperature and is important to test for possible asymmetries in polarization (see for instance 
\cite{Dvorkin:2007jp} for the description of the polarization field in models
that break statistical isotropy locally through a modulation field). 
We confirm the $TT$ anomalies that have been already reported by severals groups.
Our analysis of polarized and cross-spectra does not show significant anomalies, as from Table \ref{tableprobabilities2}.

The origin of the these hemispherical asymmetries is still unknown. They can be primordial or
due to some residual foreground or systematic effect.  For instance, in \citet{Li:2009zya} an anomalous correlation between temperature and 
observation number has been claimed to be present in the WMAP 5yr data, potentially impacting the large scale pattern of CMB maps 
(including the power asymmetries). In that paper, this effect has been related to an imbalance in the differential observation scheme of the WMAP mission.
{\sc Planck} is observing the sky with a totally different scheme and therefore is free from this particular systematic effect.

{\sc Planck} will be able to furtherly confirm the temperature anomalies and shed new light onto the polarization sector: 
the quality of {\sc Planck} data \citep{bluebook, 2010arXiv1001.3321B, 2010arXiv1001.2657M, 2010arXiv1001.4562M} is expected to be 
sufficiently high to use the $R$ estimator even for polarization and cross spectra.
We show the improvements for the $D$ estimator expected from {\sc Planck} in Fig. \ref{compPlanckWMAP}. 

No significant differences have been found by masking the Cold Spot in the Southern hemisphere with a disk of $8$ degrees of radius (conservative choice). 
Despite of the possible causes we mentioned in Section 3, we think that our results in this respect is worthy of note.
Further investigation is needed
since a correlation between these two anomalies (i.e. the Cold Spot and the North-South asymmetry) has been recently claimed \citep{Bernui:2009pv}, especially in the polarization sector where the properties of the Cold Spot are still unclear \citep{Vielva2010}.

\section*{Acknowledgements}
We thank Patricio Vielva for useful discussions on the Cold Spot.
We acknowledge the use of the BCX and SP6 at CINECA under the agreement INAF/CINECA and the use of computing facility at NERSC.
We acknowledge use of the HEALPix \citep{gorski} software and analysis package for
deriving the results in this paper.  
We acknowledge the use of the Legacy Archive for Microwave Background Data Analysis (LAMBDA). 
Support for LAMBDA is provided by the NASA Office of Space Science.
We aknowledge support by ASI through ASI/INAF Agreement I/072/09/0 for the {\sc Planck} LFI Activity of
Phase E2 and I/016/07/0 COFIS.

\end{document}